\documentclass[12pt,draftcls,onecolumn]{IEEEtran}

\IEEEoverridecommandlockouts

\usepackage{amsmath}
\usepackage{amsfonts}
\usepackage{amssymb}
\usepackage{pdfpages}
\newtheorem{theorem}{Theorem}
\newtheorem{proposition}[theorem]{Proposition}

\title{Discussion on ``AC Drive Observability Analysis''}

\author{Mohamad Koteich, \textit{Student Member, IEEE}, \\ Abdelmalek Maloum, Gilles Duc and Guillaume Sandou
	\thanks{Manuscript received February 18, 2015; revised April 3, 2015; accepted May 2, 2015.}
	\thanks{Copyright \copyright~2015 IEEE. Personal use of this material is permitted. However, permission to use this material for any other purposes must be obtained from the IEEE by sending a request to pubs-permissions@ieee.org.}
	\thanks{Mohamad Koteich is with Renault S.A.S. Technocentre, 78288 Guyancourt, France, and also with L2S - CentraleSup\'{e}lec - CNRS - Paris-Sud University, 91192 Gif-sur-Yvette, France (e-mail: mohamad.koteich@renault.com).}
	\thanks{Abdelmalek Maloum is with Renault S.A.S. Technocentre, 78288 Guyancourt, France (e-mail: abdelmalek.maloum@renault.com).}
	\thanks{Gilles Duc and Guillaume Sandou are with L2S - CentraleSup\'{e}lec - CNRS - Paris-Sud University, 91192 Gif-sur-Yvette, France  (e-mail: gilles.duc@centralesupelec.fr; guillaume.sandou@centralesupelec.fr).}}

\begin{document}
	\maketitle
	\markboth{IEEE TRANSACTIONS ON INDUSTRIAL ELECTRONICS}%
	{Koteich \MakeLowercase{\textit{et al.}}: Discussion on ``AC Drive Observability Analysis''}
	
	\begin{abstract}
		In the paper by Vaclavek \textit{et al.} (\textit{IEEE Trans. Ind. Electron.}, vol. 60, no. 8, pp. 3047-3059, Aug. 2013), the local observability of both induction machine and permanent magnet synchronous machine under motion sensorless operation is studied. In this letter, the ``slowly varying'' speed assumption is discussed, and the permanent magnet synchronous machine observability condition at standstill is revisited.
	\end{abstract}
	
	
	\section{Introduction}
	\IEEEPARstart{I}{n} the above paper \cite{vaclavek13} the local observability of the induction machine (IM) and the permanent magnet synchronous machine (PMSM) is studied. Obviously, it is a very good paper as it has been referred to by many others since it was published \cite{ref1} \cite{ref2} \cite{ref3}. To the best of our knowledge, it is the first paper that presents the PMSM observability conditions in the rotating reference frame, which provides useful explicit conditions.
	
	The observability analysis in \cite{vaclavek13} is restricted to the ``slowly varying'' speeds. This means that the obtained observability conditions for both machines are valid only under constant (or nearly constant) speed operating conditions. One can argue that this is not the case for a wide range of electrical drive applications.
	
	\section{Further Remarks on the IM Observability}
	
	It is worth mentioning that the IM observability study made by de Wit \textit{et al.} \cite{canudas00} covers both cases: 1) constant speed, which leads to the $5-$dimensional machine model adopted by \cite{vaclavek13}, and 2) varying speed under slowly varying load torque assumption, which leads to a $6-$dimensional model by adding the load torque to the state vector \cite{ref4}. The only additional parameter needed in the second case is the rotor and load inertia, which can be fairly accurately known in numerous applications. The observability condition of the $6-$dimensional model can be expressed, using the same notations as \cite{vaclavek13}, as:
	
	\begin{eqnarray}
	\frac{\xi_2}{\omega_e^2 + \xi_2^2}  \frac{d\omega_e}{dt} \mathbf{\Psi}_r^{\ast}.\mathbf{\Psi}_r - {\frac{d \mathbf{\Psi}_r}{dt}\times \mathbf{\Psi}_r}\neq 0
	\label{im_condition}
	\end{eqnarray}
	One can design an observer for the IM under slowly varying speed operating conditions. In this case, the observability condition \eqref{im_condition} becomes equivalent to the condition calculated for the $5-$dimensional model (as discussed in \cite{canudas00}). This provides a more general IM local observability analysis.
	
	\section{Comments on the PMSM Observability}
	
	The slowly varying speed assumption is not required in the PMSM observability analysis, where only the first order derivatives of the stator currents are evaluated. Thus, the PMSM observability conditions presented in \cite{vaclavek13} are valid for any rotor acceleration.
	
	The determinant of the observability matrix numbered (97) in the paper under discussion can be written as\footnote{Symbolic math software is used to reproduce the determinant expression.}:
	\begin{eqnarray}
	D &=&\frac{1}{L_d L_q} \left[\left(\Delta L i_d + K_e \right)^2 + \Delta L^2 i_q^2
	\right]\omega_e \nonumber \\
	&&+~ \frac{\Delta L}{L_d L_q} \left[
	\Delta L\frac{di_d}{dt} i_q - 
	\left(\Delta L i_d + K_e \right)  \frac{di_q}{dt}
	\right]
	\end{eqnarray}
	In their analysis of the above equation, the authors in \cite{vaclavek13} formulate the following observability condition at standstill:
	\begin{eqnarray}
	\left|i_d + \frac{K_e}{\Delta L}\right| \left|C\right| \neq \left|i_q\right|
	\label{cc}
	\end{eqnarray}
	They claim that it is not necessary to determine the value of the constant $C$, and that in the zero or low-speed region ``the rotor position will be observable if stator current components in rotating reference frame $i_d$, $i_q$ are changing and not kept to be linearly dependent. Stator current space vector should change not only its magnitude but also direction in the rotating reference frame''.
	
	The above conclusion is unclear and yet inaccurate. First of all, the conclusion should not only concern the rotor position, the rotor angular speed observability should also be concerned. Even though it seems to be intuitive that the loss of observability concerns rather the position, this cannot be proved unless a detailed study of the \textit{indistinguishable dynamics} is done, similarly to the study done in \cite{rojas} for induction machines.
	In addition, the stator current space vector can change both its magnitude and direction without ensuring the motor observability at standstill, as shown in the sequel. 
	
	The observability condition $D \neq 0$ can be written as:
	\begin{eqnarray}
	\omega_e \neq \frac{\left(\Delta L i_d + K_e \right) \Delta L \frac{di_q}{dt} - \Delta L \frac{di_d}{dt} \Delta L i_q}{\left(\Delta L i_d + K_e \right)^2 + \Delta L^2 i_q^2}
	\end{eqnarray}
	which gives:
	\begin{eqnarray}
	\omega_e \neq   \frac{d}{dt} \arctan \left(\frac{\Delta L i_q}{\Delta L i_d + K_e}\right)
	\label{cond_w}
	\end{eqnarray}
	
	Let's define a fictitious \textit{observability vector} $\Psi_\mathcal{O} = \Psi_{\mathcal{O}d} + j \Psi_{\mathcal{O}q}$ which has the following components in the rotating ($dq$) reference frame:
	\begin{eqnarray}
	\Psi_{\mathcal{O}d} &=& \Delta L i_d + K_e\\
	\Psi_{\mathcal{O}q} &=& \Delta L i_q
	\end{eqnarray}
	
\begin{figure}[!t]
		\centering
		\includegraphics[scale=1]{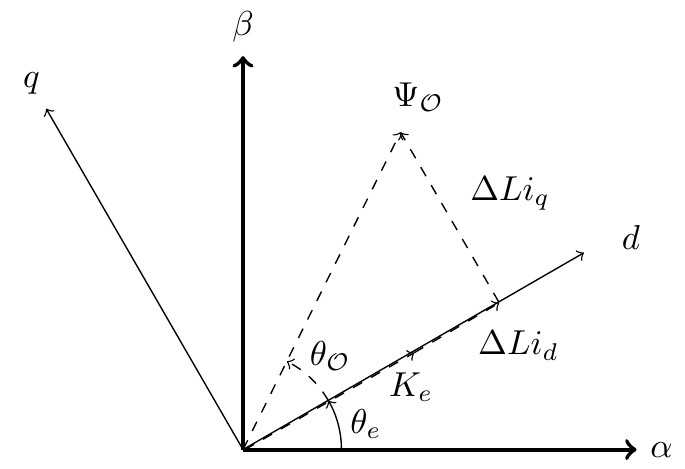}
		\caption{Vector diagram of the fictitious observability vector (dashed).}
		\label{obsv_vector}
	\end{figure}

	Then, the condition \eqref{cond_w} can be formulated as:
	\begin{eqnarray}
	\omega_e \neq   \frac{d}{dt} \theta_\mathcal{O}
	\label{condition}
	\end{eqnarray}
	where $\theta_\mathcal{O}$ is the phase of the vector $\Psi_\mathcal{O}$ in the rotating reference frame (Figure \ref{obsv_vector}). The following sufficient condition for the PMSM local observability can be stated:
	\begin{proposition}
		The PMSM is locally observable if the angular speed of the fictitious vector $\Psi_\mathcal{O}$ in the $dq$ reference frame is different from the rotor electrical angular speed in the stator reference frame. 
	\end{proposition} 
	
	At standstill, the above condition becomes: the vector $\Psi_\mathcal{O}$ should change its orientation in order to ensure the local observability. This provides a better formulation of the PMSM observability conditions.
	
	It turns out that the $d-$axis component of the vector $\Psi_\mathcal{O}$ is nothing but the so-called ``active flux'' introduced by Boldea \textit{et al.} in \cite{boldea08} (also called ``fictitious flux'' by Koonlaboon \textit{et al.} \cite{koonlaboon05}), which is, by definition, the torque producing flux aligned to the rotor $d-$axis. The $q-$axis component of the vector $\Psi_\mathcal{O}$ is related to the saliency ($\Delta L$) of the machine, and is aligned to the rotor $q-$axis.
	
\begin{figure}[!t]
		\centering
		\includegraphics[scale=1.2]{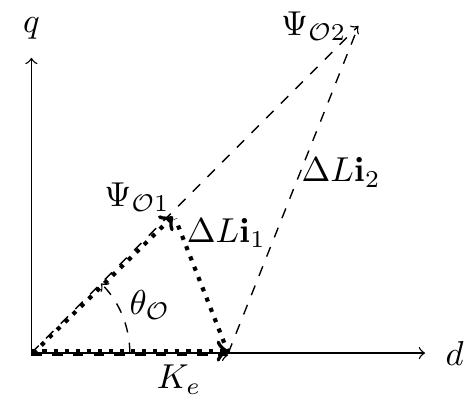}
		\caption{Vector diagram illustrating two observability vectors that correspond to two stator current space vectors that differ in magnitude and direction.}
		\label{obsv_vector2}
	\end{figure}

	Figure \ref{obsv_vector2} shows two observability vectors, $\Psi_{\mathcal{O}1}$ (dotted) and $\Psi_{\mathcal{O}2}$ (dashed), that correspond to two different (in magnitude and direction) current space vectors:
	\begin{eqnarray}
	\mathbf{i_1} &=& i_{d1} + j i_{q1} ~~;~~ (i_{d1} < 0 ~,~ i_{q1} > 0)\\
	\mathbf{i_2} &=& i_{d2} + j i_{q2} ~~;~~ (i_{d2} > 0 ~,~ i_{q2} > 0)
	\end{eqnarray}
	It is obvious that, contrary to the conclusion drawn in \cite{vaclavek13} for the IPMSM, at standstill ($\omega_e = 0$), the stator current space vector can change both its magnitude and direction, following the constant-$\theta_{\mathcal{O}}$ trajectory, without fulfilling the observability condition \eqref{condition}. It should be noticed that this is related to the constant $C$ of the equation \eqref{cc}, which value is judged to be unnecessary in the paper under discussion.
	
	As for the surface-mounted (S) PMSM ($\Delta L = 0$) under sensorless operation, the fictitious observability vector is equal to the rotor permanent magnet flux vector, which is fixed in the $dq$ reference frame. This means that the observability problem arises only at standstill, which is consistent with the conclusion on SPMSM observability in the discussed paper \cite{vaclavek13}.
	
	\section*{Erratum}
	In the list of references of \cite{vaclavek13}, the reference number 17 is not correctly cited; the name of the first author is omitted. The correct citation is \cite{chiasson05}.
	
	\bibliographystyle{ieeetr}
	\bibliography{BIBLIO_15-TIE-0524}
	
\end{document}